\documentstyle[epsfig,art12]{article}
\def\p2t{p_{2\bot}}
\def\q2t{q_{2\bot}}
\begin{document}
\begin{center}
  \begin{Large}
  \begin{bf}
BFKL gluon dynamics in the $p_T(D^*)$ spectra at HERA\\
%
  \end{bf}
  \end{Large}
  \vspace*{5mm}
  \begin{large}
    S.P. Baranov\\
  \end{large}
P.N. Lebedev Physical Institute, Russian Academy of Science\\  117924
Moscow, Russia\\
  \vspace{5mm}
  \begin{large}
    N.P. Zotov\\
  \end{large}
D.V. Skobeltsyn Institute of Nuclear Physics, \\
M.V. Lomonosov Moscow State University,
    119899 Moscow, Russia\\
\end{center}
  \vspace{10mm}
\begin{abstract}
\noindent
In the framework of semihard QCD approach, with the emphasis on the BFKL 
evolution of gluon distributions,
we calculate the differential cross sections of inclusive $D^{*\pm}$
mesons electroproduction at HERA and confront our results with  available
ZEUS experimental data.
\end{abstract}

\section{Introduction}

Recently H1 and ZEUS Collaborations  have reported~\cite{r1,r2} experimental data on the differential
cross section $d^2\sigma/d\eta dp_{T}$ of inclusive $D^{*\pm}$ electroproduction at low $Q^2$ 
(equivalent of photoproduction). These results were compared with NLO pQCD
calculations in the so called the massive  and massless charm schemes~\cite{r3,r4}.
At the DIS 98 Workshop M. Sutton has expressed~\cite{r5} the ZEUS Collaboration point 
of view by the statement that the calculations underestimate the cross section in the 
intermediate $p_{T}(D^*)$ and forward $\eta (D^*)$ regions.
 
The massive scheme meets similar difficulties at Tevatron conditions also.
To reproduce the heavy quarks $p_T$ spectra one usually introduces the primordial 
$k_T$ of the incoming partons. The size of this $k_T$ cannot be predicted within 
the model itself and is required to be of about 1 or 2 GeV to fit the data.
The massless scheme is only valid in the asymptotic limit $p_T \gg m$ and is not
suitable for intermediate $p_T(D^*)$ region at HERA~\cite{r6}. Also, it uses 
some special assumptions on the $c\to D^*$ fragmentation and needs unusually low
value of the Peterson parameter $\epsilon\simeq 0.02$~\cite{r7}.
In view of the above problems, it would be certainly reasonable to try a different way.

In the present note, we focus on the so called semihard approach~\cite{r8} (SHA),  
which we had applied earlier~\cite{r9} to open charm and $J/\Psi$ photoproduction. 
The inherent BFKL gluon evolution dynamics atomatically generates $k_T$ distributions
and thus allows to keep this theoretical input under control.


 At the HERA energies and beyond, the interaction dynamics is governed
 by the properties of parton distributions in the small $x$ region.
 This domain is characterized by the double inequality
 $s\gg\mu^2\simeq\hat s\gg\Lambda^2$,
 which shows that the typical parton interaction scale $\mu$ is much higher
 than the QCD parameter $\Lambda$, but is much lower than the total c.m.s.
 energy $\sqrt s$. The situation is therefore classified as ``semihard''.
 


The resummation~\cite{r8},\cite{r10}-\cite{r12} of the terms
 $[\mbox{ln}(\mu^2/\Lambda^2)\,\alpha_s]^n$,\\
 $[\mbox{ln}(\mu^2/\Lambda^2)\,\mbox{ln}(1/x)\,\alpha_s]^n$ and
 $[\mbox{ln}(1/x)\,\alpha_s]^n$ in the SHA 
 results in the unintegrated parton distributions
 $\varphi_i(x,k_{T}^2,\mu^2)$.  They obey the BFKL \cite{r13}
 equation and reduce to the conventional
 parton densities once the $k_{T}$ dependence is integrated out:
\begin{equation} \label{kt}
\int_0^{\mu^2}\!\!\varphi_i(x,k_{T}^2,\mu^2)\;dk_{T}^2=x\,F_i(x,\mu^2).
\end{equation}

The specific properties of semihard theory may manifest in several ways.
With respect to inclusive production properties, one points out an
 additional contribution to the cross sections due to the integration over
 the $k_{T}^2$ region above $\mu^2$ and the broadening of the $p_T$ spectra
 due to extra transverse momentum of the interacting gluons~\cite{r8}-\cite{r10}.
It is important that the gluon is not on mass shell
 but is characterized by virtual mass proportional to its transverse
 momentum: $m^2=-k_{T}^2/(1-x)$. This also assumes a modification of the
 polarization density matrix. One striking consequence on the $J/\psi$ spin
alignement has been demonstrated in~\cite{r14}. 
Now consider the SHA predictions for $D^*$ photoproduction at HERA.

\section{$D^{*\pm}$ electroproduction in the QCD semihard approach}

The differential cross section of the process $e(p_e) + g(k) \to  e'(p'_e) +
c(p_1) + \bar c(p_2)$ is calculated exactly. It has the form
\begin{equation}
d\sigma \sim L^{(e)}_{\mu\mu'} H^{\mu\nu}\bar H_{\mu'\nu'} L^{(g)}_{\nu\nu'}
G(x, k^2_{T}) dp^2_{1T}dp^2_{2T}dp'^2_{eT}dy_1dy_2.
\end{equation}

When calculating the spin average of the matrix element squared,
we substitute the full lepton tensor for the photon polarization matrix:
\begin{equation}
\overline{\epsilon_1^{\mu}\epsilon_1^{*\nu}}=
4\pi\alpha [8p_e^{\mu}p_e^{\nu}-4(p_ek_1)g^{\mu\nu}]/(k_1^2)^2  \label{epsgam}
\end{equation}
(including also the photon propagator factor and photon-lepton coupling).

 The virtual gluon polarization tensor is taken~\cite{r8}:
\begin{equation}
L^{(g)}_{\mu\nu} \sim  \overline{\epsilon^{\mu}\epsilon^{*\nu}}
  =k_{T}^\mu k_{T}^\nu/|k_{T}|^2, \quad \label{epsk}
\end{equation}
 The above prescription has a clear analogy with the equivalent
 photon approximation.
 Neglecting the second term in (\ref{epsgam}) the right hand side in the small 
 $x$ limit, $p\gg k$, one immediately arrives at the spin structure
 $\overline{\epsilon^{\mu}\epsilon^{*\nu}}\sim L^{\mu\nu} \sim p^\mu p^\nu$.
 The latter may be rewritten in the form (\ref{epsk}) if to use the
 parametrization for the 4-vectors $k=xp+k_{T}$
 and to apply a gauge shift $\epsilon^\mu \to \epsilon^\mu - k^\mu/x$.

Concerning another essential ingredient, the unintegrated gluon distribution
$\varphi_G(x,k^2_{T},\mu^2)$ we follow the prescriptios of paper~\cite{r15}. The proposed 
method lies upon a straightforward perturbative solution of the
BFKL equation where the collinear gluon density $x\,G(x,\mu^2)$ is used as
the boundary condition in the integral form (\ref{kt}):

Technically, the unintegrated gluon density is calculated as a convolution
of collinear gluon density with universal weight factors \cite{r15}:
\begin{equation} \label{conv}
 \varphi_G(x,k_{\perp}^2,\mu^2) = \int_x^1
 {\cal G}(\eta,k_{\perp}^2,\mu^2)\,
 \frac{x}{\eta}\,G(\frac{x}{\eta},\mu^2)\,d\eta,
\end{equation}
\begin{equation} \label{J0}
 {\cal G}(\eta,k_{\perp}^2,\mu^2)=\frac{\bar{\alpha}_s}{xk_{\perp}^2}\,
 J_0(2\sqrt{\bar{\alpha}_s\ln(1/\eta)\ln(\mu^2/k_{\perp}^2)}),
 \qquad k_{\perp}^2<\mu^2,
\end{equation}
\begin{equation}\label{I0}
 {\cal G}(\eta,k_{\perp}^2,\mu^2)=\frac{\bar{\alpha}_s}{xk_{\perp}^2}\,
 I_0(2\sqrt{\bar{\alpha}_s\ln(1/\eta)\ln(k_{\perp}^2/\mu^2)}),
 \qquad k_{\perp}^2>\mu^2,  
\end{equation}
where $J_0$ and $I_0$ stand for Bessel functions (of real and imaginary
arguments, respectively), and $\bar{\alpha}_s=\alpha_s/3\pi$.

 This provides a common smooth parametrization for the whole
$k_{T}$ range without  extra free parameters. For the input 
collinear densities we use the standard GRV set~\cite{r16} that
fixes the absolute normalization of $\varphi_G$.

Finally the fragmentation $c \to D^*$ is described by Peterson fragmentation
function using usual value of parameter $\epsilon  \,= \,0.06$~\cite{r17}.
We used in our calculations for charm quark mass the value $m_c \,= \,1.3$ GeV
or $m_c \,= \,1.5$ GeV and $\mu^2 \,= \,m^2_T$.

\section{Results and discussions}

The theoretical results obtained in SHA for $p_{T}(D^*)$ and 
$\eta(D^*)$ distributions are plotted in Figs. 1 and 2,
respectively, together with ZEUS experimental data collected
 in  kinematic region~\cite{r2}:
$Q^2 < 4 \mbox{GeV}^2$, -1.5  $\eta < 1$,  $115 <  W < 280 \mbox{GeV}$~\cite{r2}
in Figs 1 and 2.

We see that semihard approach reproduces the experimental $p_{T}$ distribution.
As is expected, it is broader than pQCD massive and massless schemes results.  
Although the shape of the gluon $k_T$ distribution depends on the parameter $\mu$ 
in the collinear gluon density $x\,G(x,\mu^2)$, its evolution is under control by BFKL
equation. This contrasts with the massive and massless schemes where the gluon $k_T$
value is taken an arbitrary parameter.

Concerning the pseudorapidity distributions (Fig.2), semihard approach is also
in a satisfactory agreement with data. The agreement looks even better than in the
pQCD massless scheme, although the latter has spent special efforts 
(such as the modification of charm fragmentation function).

In general our numerical results are the closer to massless scheme than to massive one.
According to the investigations by B.A. Kniehl et al.~\cite{r6} the massless
cross section corresponds to LO + NLO + NNLO of the direct photon contribution
in the massive cross section. SHA, even with the LO matrix elements only, efficiency takes
into account higher orders contributions by the BFKL evolution of gluon densities.
The structure of these terms is different from those in the massless scheme, but their
numerical effect appears to be rather similar.

\section{ Acknowledgements}
We thank  L. Gladilin for providing us with ZEUS experimental data.

\newpage

\begin{figure}[t]
    \hspace*{-0.5cm}
    \mbox{\epsfig{file=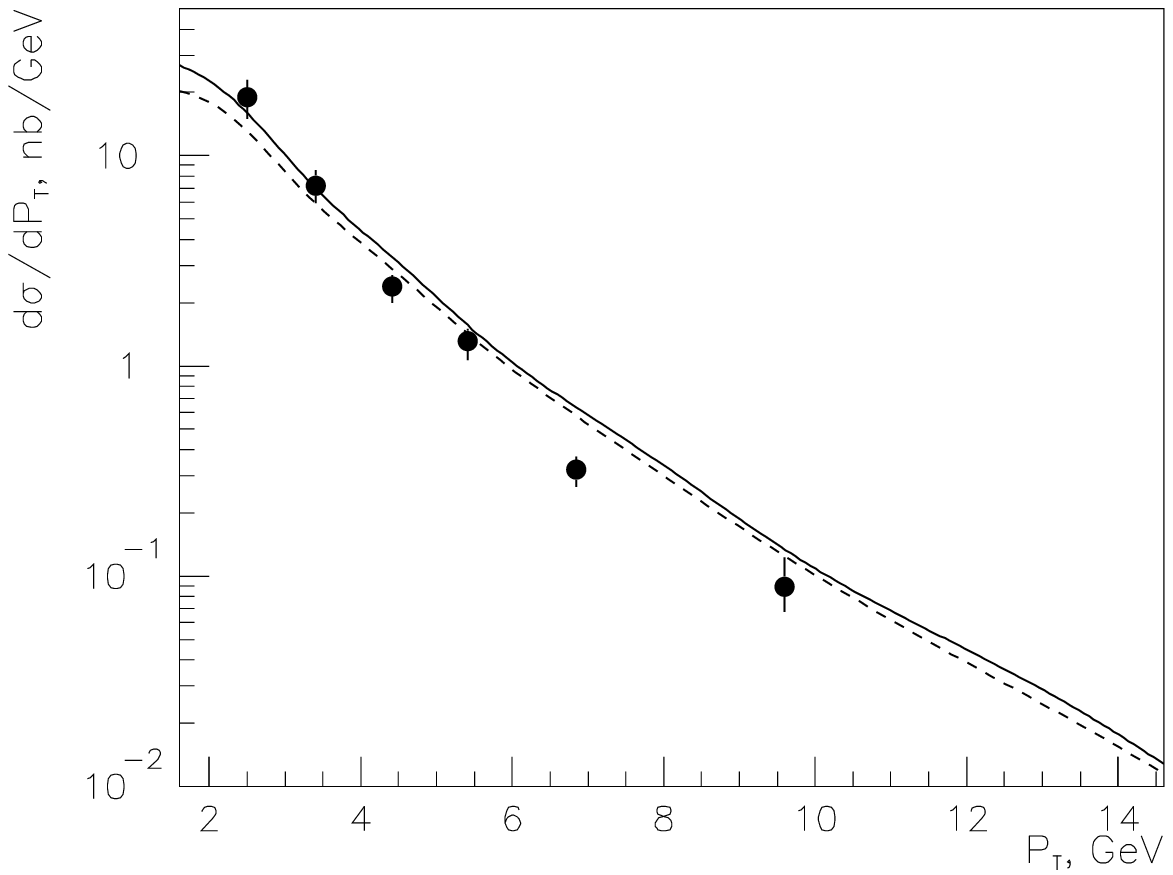,height=7.5cm}}
    \caption{{
    Differential cross section $d\sigma/dp_T$ 
    for $D^{*\pm}$ electroproduction at HERA. 
    Solid curve: $m_c=1.3$~GeV, dashed curve: $m_c=1.5$~GeV.}}
\end{figure}

\begin{figure}[b]
    \hspace*{0.3cm}
    \mbox{\epsfig{file=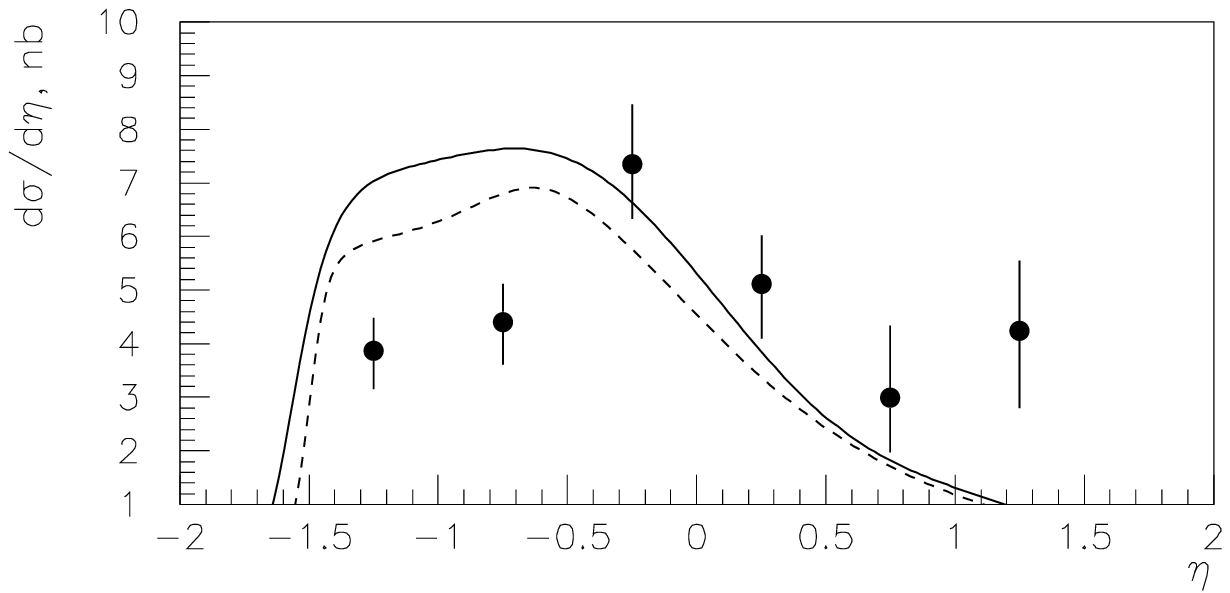,height=4.5cm}}
    \caption{{
    Differential cross section $d\sigma/d\eta$
    for $D^{*\pm}$ electroproduction at HERA.
    Solid curve: $m_c=1.3$~GeV, dashed curve: $m_c=1.5$~GeV.}}
\end{figure} 

\end{document}